\documentclass[dvips]{amsart}
\usepackage{amsmath}%
\usepackage{amsfonts}%
\usepackage{amssymb}%
\usepackage{graphicx}
\newtheorem{theorem}{Theorem}
\theoremstyle{plain}

\newtheorem{algorithm}{Algorithm}
\newtheorem{axiom}{Axiom}
\newtheorem{case}{Case}
\newtheorem{claim}{Claim}
\newtheorem{conclusion}{Conclusion}
\newtheorem{condition}{Condition}
\newtheorem{conjecture}{Conjecture}
\newtheorem{corollary}{Corollary}
\newtheorem{criterion}{Criterion}
\newtheorem{definition}{Definition}
\newtheorem{example}{Example}
\newtheorem{exercise}{Exercise}
\newtheorem{lemma}{Lemma}
\newtheorem{notation}{Notation}
\newtheorem{problem}{Problem}
\newtheorem{proposition}{Proposition}
\newtheorem{remark}{Remark}
\newtheorem{solution}{Solution}
\newtheorem{summary}{Summary}
\numberwithin{equation}{section}
\begin{document}
\title{BRST invariance in Coulomb gauge QCD}
\author{A. Andra\v si}
\address[A. Andra\v si]{Vla\v ska 58, Zagreb, Croatia}
\author{J C Taylor}
\address[J C Taylor]{DAMTP, University of Cambridge, Cambridge, UK}

\date{July 1, 2015}


\begin{abstract}
In the Coulomb gauge, the Hamiltonian of QCD contains terms of order $\hbar^2$, identified by Christ and Lee, which are non-local
but instantaneous. The question is addressed how do these terms fit in with  BRST invariance. Our discussion is confined to the simplest,
$O(g^4)$, example.
 
\end{abstract}
\maketitle

\section{Introduction}

In the Feynman gauge, Lorentz invariance is obvious, but unitarity requires BRST invariance for its justification.
In the Coulomb gauge, on the other hand, unitarity is explicit, and Lorentz invariance is a consequence of the commutation relations
between the generators of the Lorentz group \cite{sch}. There seems to be no role for BRST invariance. That would be the case
if the theory were finite (in perturbation theory). However, there are two sorts of divergent integrals in the Coulomb gauge: ordinary UV divergences,
and \textit{energy divergences}, that is integrals which are divergent over the energy integrals with the spatial momenta held fixed.
Each of these separately seems to necessitate consideration of BRST invariance.
 
First, the renormalization of the UV divergences requires counter-terms which do not appear in the original Hamiltonian. These require BRST
invariance to control them (see for example section 12.4 of \cite{iz} or chapter 17 of \cite{w}). The renormalized Hamiltonian can be got into the same form as the bare one by field redefinitions, involving for example
mixing of the electric field $E_i^a$(the momenta) with $f^a_{0i} \equiv \partial_0 A_i^a-\partial_i A_0^a+gf^{abc}A_0^b A_i^c$ \cite{aajct}.

Second, there are complications arising from energy divergences like
\begin{equation}
\int dp_0dq_0 \frac{p_0}{p_0^2-P^2+i\epsilon}\frac{q_0}{q_0^2-Q^2+i\epsilon}
\end{equation}
which occur at 2-loop order
($P$, $Q$ denote magnitudes of spatial momenta). Before discussing such integrals, we review the work of Christ and Lee and others \cite{cl}
on the operator ordering of the Coulomb gauge Hamiltonian. This  Hamiltonian contains a Coulomb potential term, $C(\hat{A}_i^a,\hat{E}_j^b)$ which is
a non-local but instantaneous functional of the vector potential $\hat{A}_i^a$ and the electric field $\hat{E}_j^b$ ($i,j...$ are spatial indices and $a,b,..$ are colour
indices). According to \cite{cl}, if the Hamiltonian is Weyl ordered, extra terms appear in the Hamiltonian, called $V_1$ and $V_2$ which are of order
$\hbar^2$ and are also non-local but instantaneous functionals of $\hat{A} _i^a$. It is also argued in \cite{cl} that the Feynman rules derived from  a Weyl ordered
Hamiltonian do not generate integrals like (1). In this treatment, the operator $\hat{A}_i^a$ is, by definition, divergenceless.

But there is another approach, which is not to Weyl order the Hamiltonian, and to allow integrals like (1) to arise in Feynman diagrams, but seek to combine graphs 
so that the combinations are free of energy divergences. This was done in \cite{doust}. The Feynman integrals allow one to calculate the effective
action $\Gamma(A_i^a,E_j^b,A_0^c)$. In this effective action, $A_i^a$ need not be divergenceless. The theorem of \cite{doust} was extended by us
\cite{aajct2} to remove the restriction that $A_i^a$ be divergenceless, thus effectively giving generalizations $\bar{V}_1(A_i^a)+\bar{V}_2(A_i^a)$ of the Christ-Lee
functionals when $A_i^a$ is not necessarily divergencless.

In this second approach, it is necessary temporarily, while graphs are combined, to regulate integrals like (1). Dimensional regularization, which we use to control the UV divergences, is useless for energy divergences. A convenient method of regularization (and the only one we know) is to use a flow, or interpolating, gauge.
A simple choice is defined by the gauge-fixing term
\begin{equation}
 \frac{1}{2\theta^2}(\partial^iA_i^a-\theta^2 \partial^0 A_0^a)^2
\end{equation}
where $\theta$ is a parameter. The Coulomb gauge is then defined by the limit $\theta \rightarrow 0$ (and the Feynman gauge is $\theta=1$, but this is not relevant here). In most graphs, the limit can be taken before
integrating; but, in graphs involving integral like (1), the energy integral must be done first.

In this approach, BRST invariance is necessary in order to guarantee that S-matrix elements are independent
of $\theta$, and to guarantee the Lorentz invariance when $\theta\neq 0$.

The BRST identities in general involve
\begin{equation}
 \frac{d}{dx^i}\left[\frac{\delta \Gamma}{\delta A_i^a(x)}\right]
\end{equation}
and so it is necessary that all the contributions to $\Gamma$ must be defined for fields $A_i^a$ which are not
divergenceless. This means that the Christ-Lee functions need to be generalized, as was done in \cite{aajct2}.
The purpose of this paper is to identify the role that the (generalized) Christ-Lee functions play in a BRST identity.

\section{The BRST identity and Christ-Lee functions}

We will use the following notation for vectors: $p_{\mu}=(p_0,\textbf{P})=(p_0,P_i)$ with $i=1,...,n$
and $P=|\textbf{P}|$.

The generalized Christ-Lee functions $\bar{V}_1+\bar{V}_2$ are instantaneous (that is, independent of
energies). We consider as an example the two-point  ($O(g^4)$) terms. They contain the time delta function
\begin{equation}
\delta(t_1-t_2).
\end{equation}

There is a BRST identity
\begin{equation}
k_0 \Pi_{0j}(k)+K_i\Pi_{ij}(k)=[K^2\delta_{{j'j}}-K_jK_{j'}]Q_{j'}(k)
\end{equation}
where $Q_{j'}(k)$ is a contribution coming from the ghost dependence of $\Gamma$. In this example, $Q_{j'}$ must be
proportional to $K_{j'}$, and so the right hand side of (2.2) vanishes, and the BRST identity looks just like
a Ward identity. Since each term must be proportional to $K_j$, we lose nothing by contracting with $K_j$,
and this is what, for simplicity, we do, obtaining
\begin{equation}
k_0K_j\Pi_{0j}+K_iK_j\Pi_{ij}=0.
\end{equation}

The Christ-Lee term is a contribution to $\Pi_{ij}$ which is independent  of $k_0$ and so in the Fourier transform has $\delta(t_1-t_2)$. This requires $\Pi_{0j}$ to have contribution proportional to $\epsilon(t_1-t_2)$. We show that there are graphs which give the required contribution. Graphs (a) and (b) in Fig.1 each contain
the integral

\begin{equation}
\int dp_0 dq_0 dr_0 \delta(p_0+q_0+r_0-k_0)\frac{p_0}{p^2+i\epsilon}\frac{q_0}{q^2+i\epsilon}\frac{r_0}{r^2+i\epsilon}$$
$$=-\pi^2 \frac{k_0}{k_0^2-(P+Q+R)^2+i\epsilon}
\end{equation}
 The Fourier transform of (2.4) is
\begin{equation}
-\pi^2\epsilon(t)\exp[-i|t|(P+Q+R-i\epsilon)]
\end{equation}
containing  the required $\epsilon (t)$ step function.

\section{A simple example}
\begin{figure}[h]
\centering
\includegraphics[width=0.8\textwidth]{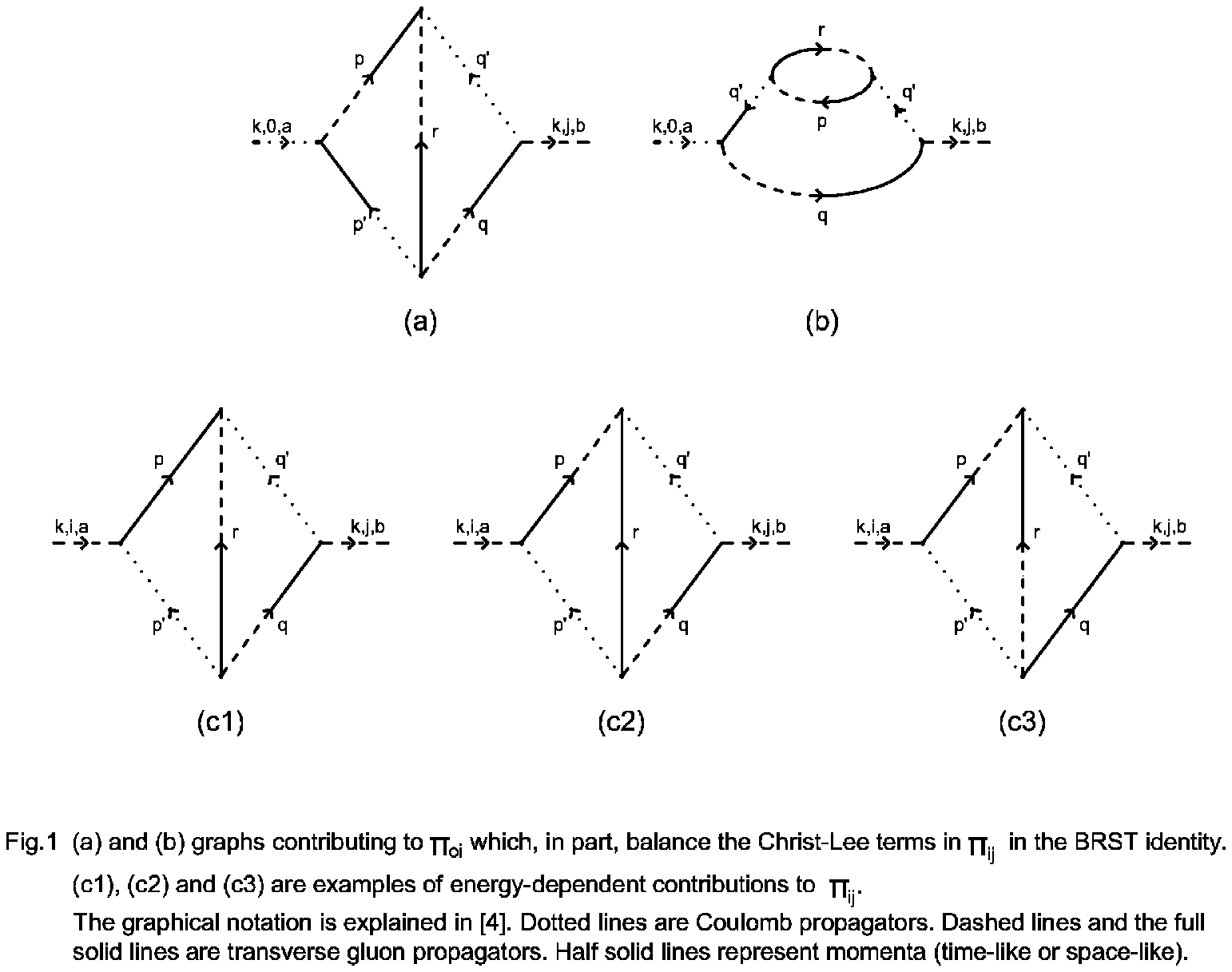}
\end{figure}
This example involves the $\Pi_{0i}$ graph (a) and the $\Pi_{ij}$ graphs (c1), (c2), (c3), together with the three Doust graphs
like (A1) in Fig.2 of \cite{aajct3}. Graph (a) involves the integral (2.4). The graph (c1)
has the energy integral
\begin{equation}
\int dp_0 dq_0 dr_0 \delta(p_0+q_0+r_0-k_0)\frac{P^2}{p^2}\frac{1}{q^2}\frac{1}{r^2}
 =- \pi^2\frac{P(P+Q+R)}{k_0^2-(P+Q+R)^2}.
\end{equation}
Graphs (c2) and (c3) are obtained from (3.1) by permuting $P,Q,R$, so the total is
\begin{equation}
-\pi^2 \frac{(P+Q+R)^2}{k_0^2-(P+Q+R)^2}.
\end{equation}

When graphs (a) and (c) are inserted into the identity (2.3), expression (2.4) is multiplied by $k_0$ and the remaining parts of the integrands are identical except for a minus sign (using the identity $K_iT_{il}(P)=-P'_iT_{il}(P)$,
$T_{il}$ being the transverse projection operator and $P'_i=P_i-K_i$). Thus (2.4) and (3.2) combine to give
the $k_0$-independent result $-\pi^2$. Most of this result  is exactly what comes from the sum of the three Doust graphs like (A1) of \cite{aajct3}, each of which has coefficient $-\pi^2/3$. The exception is the contribution from the $\delta_{lm}$ term in the projection operator
\begin{equation}
T_{lm}(R)\equiv \delta_{lm}-\frac{R_lR_m}{R^2}
\end{equation}
 on the $r$-line. This contributes
\begin{equation}
-c\delta_{lm}\pi^2\int d^nP \frac{K.P'P_l}{P^2P'^2}\int d^nQ \frac{Q'.KQ_m}{Q^2Q'^2}=-\frac{1}{16}c\pi^2K^6\int d^nP\frac{1}{P^2P'^2}\int d^nQ\frac{1}{Q^2Q'^2}
\end{equation}
(the coefficient $c$ is defined below in (4.5)). We return to this contribution in the next section.

This example shows that energy-independent Christ-Lee terms are essential for the BRST identity
to be satisfied. Note that the energy integrals in (2.4) and (3.1) are convergent, whereas the Doust graphs
individually contain the divergent integral (1.1).

The role of graph (b) in the BRST identity is not as simple as that of (a) in the above example; so in the next section we verify only that the $k_0$-independent part of (b) satisfies the BRST identity with the relevant
Doust graphs.

\section{Other Doust graphs in BRST identities}
In this section, we show how graph (b) requires Christ-Lee terms in order to satisfy the BRST identity (2.3).
Because of (2.4),  graph (b) (like graph (a)) has a contribution to (2.3) of the form
\begin{equation}
-\pi^2
\int d^nPd^nQd^nR\delta^n(\textbf{P}+\textbf{Q}+\textbf{R}-\textbf{K}) F(P,Q,R,K)\frac{k_0^2}{k_0^2-(P+Q+R)^2}$$
$$=M(K)+N(K,k_0)
\end{equation}
where
\begin{equation}
M=-\pi^2\int d^nPd^nQd^nR\delta^n(\textbf{P}+\textbf{Q}+\textbf{R}-\textbf{K}) F(P,Q,R,K)
\end{equation}
and 
\begin{equation}
N=-\pi^2\int d^nPd^nQd^nR\delta^n(\textbf{P}+\textbf{Q}+\textbf{R}-\textbf{K}) F(P,Q,R,K)\left[\frac{(P+Q+R)^2}{k_0^2-(P+Q+R)^2}\right].
\end{equation}
The function $F$ here is quite complicated (it depends upon $P'=|\textbf{P}-\textbf{K}|$ etc. as well as on $P,Q,R$), but in (4.2) 
it can be simplified a lot (using the rules of dimensional regularization, and symmetry under $\textbf{P,Q,R}\rightarrow -\textbf{P}',-\textbf{Q}',-\textbf{R}$) to give
\begin{equation}
M=-\frac{c}{2}\int d^nPd^nQd^nR\delta^n(\textbf{P}+\textbf{Q}+\textbf{R}-\textbf{K}) \frac{(\textbf{K.P})^2}{P^2Q^2R^2}.
\end{equation}
Here  
\begin{equation}
c=(g^4/2)C_G^2(2\pi)^{-n-1},
\end{equation}
 ($C_G$ is the colour group Casimir).
 Expression (4.4)
coincides (up to a minus sign) with the Christ-Lee terms in $\Pi_{ij}$ as in  equations (5.2) coming from the B and C in \cite{aajct3}. Another term, in (5.8) of \cite{aajct3}, coming from the B graphs, coincides with the term (3.4), left over from section 3 above. Thus the BRST identity is satisfied as far as the energy-independent term $M$ in (4.2) goes.
We have not identified in a simple way how the graphs similar to (c) in Fig.1 balance the term $N$ in (4.3).

In contrast to Christ-Lee graphs, the last, $k_0$-dependent term in the integrand in (4.3) appears to tend to zero as
$k_0^2 \rightarrow \infty$, but the behaviour of the integral is not obvious. The contribution from the (c) graphs in Fig.1 (before contracting with $K_iK_j$) has the form
\begin{equation}
\delta_{ij}k_0^2f_1(K^2,k_0^2)+[K_iK_j-(K^2\delta_{ij})/n] f_2(K^2,k_0^2)+[K^2\delta_{ij}/n]f_3(K^2,k_0^2).
\end{equation}
In dimensional regularization, each of the three $f$ functions in (4.6) has dimensions of energy $(2n-6)$.
\textit{If} $f_2$ and $f_3$ are finite at $K=0$, it follows that they decrease with energy as $(k_0^2)^{n-3}$.
From  examples we have studied, we conjecture that this is the case. In this sense, the energy-dependent terms contrast with the Christ-Lee terms, which of course do not decrease with energy.

An example of a term coming into the calculation of graph (c1) is
\begin{equation}
C_G^2\frac{g^4}{2}(2\pi)^{-n-1}\int d^{n+1}Pd^{n+1}Qd^{n+1}R\\ \delta^{n+1}(p+q+r-k)\frac{1}{Q^2}\frac{1}{r^2}\frac{p_0}{p^2}\frac{q_0}{q^2}.
\end{equation}
It may be shown that this contributes to $f_3$ in (4.6) a function

\begin{equation}
\frac{c}{8}\pi^{[(9/2)-\epsilon]}2^{\epsilon}\exp(-i\pi\epsilon/4)\Gamma(\epsilon/2)\Gamma(\epsilon-1)\frac{\Gamma[1-(\epsilon/2)]}{\Gamma[(3-\epsilon)/2]}$$
$$\times \int _0^1 dr r^{-(1+\epsilon)/2} \int _0^1 dv (1-v)^{-\epsilon/2}\{1-(1-v)r\}
\left[K^2\{1-(1-v)r\}-k_0^2(1+v)\right]^{-\epsilon},
\end{equation}
which does have the above mentioned property.

\section{Summary}
We have verified that the instantaneous Christ-Lee terms contribute to the BRST identity for the two-point functions, and identified the graphs ((a) and (b) of Fig.1) with which they are linked.

\end{document}